# New ultrasmall iron-oxide nanoparticles with high magnetisation as potential $T_1$-MRI contrast agents for Molecular Imaging.

[a]*Institut de Ciència de Materials de Barcelona (ICMAB-CSIC), Esfera de la UAB, 08193 Bellaterra, Catalunya, Spain*

[b]*Department of General, Organic and Biomedical Chemistry, NMR and Molecular Imaging Laboratory, University of Mons-Hainaut, B-7000 Mons, Belgium*

* corresponding author: roig@icmab.es

## Abstract

Here we report on the synthesis of very small $\gamma$-$Fe_2O_3$ nanoparticles (5 nm) presenting very narrow particle size distribution and exceptionally high saturation magnetisation. The synthesis has been carried out in an organic medium with subsequent transfer to an aqueous solution at physiological pH. The structural and magnetic properties were kept unaltered after the solvent exchange. NMR relaxometric measurements show the potential of these particles as specific reporters for magnetic resonance molecular imaging.

## Introduction

Magnetic resonance imaging (MRI) is a medical technique widely employed in many clinical practices due to its capability to enhance contrast differences between healthy and pathological tissues. Images of body sections precisely reflect the variation in the proton density, longitudinal or transversal relaxation time, $T_1$ or $T_2$ of the tissues. Despite the inherent versatility of this imaging modality, researchers and clinicians are dedicating huge efforts to develop safer and more effective contrast agents (CAs) that will expand the diagnostic utility and improve the precision of MRI. The main application of CAs relies on the shortening of the relaxation times of the water protons. Positive contrast agents reduce $T_1$ resulting in a brighter signal, while negative contrast agents reduce $T_2$ resulting in a darker signal. The reciprocals of the relaxation times are called the relaxation rates, $R_1$ and $R_2$, with the effectiveness of a CA expressed as relaxivities, $r_1$ and $r_{2,}$ i.e. the rate(s) enhancement(s) brought per mM of metal.

A large number of compounds, mostly paramagnetic substances, have been assayed as potential MRI contrast agents[1,2]. Gadolinium chelates, such as Gd-DTPA, constitute the largest group of paramagnetic MR $T_1$-contrast agents and are considered to be safe and effective[3-5]. Iron oxides (magnetite and maghemite) play an important role as MR $T_2$-contrast agents due to their sizes and magnetic properties. Various iron oxide nanoparticles have been synthesised and evaluated as CAs[6-9]. They mainly differ in the iron oxide phase, the magnetic core size or in the type of coating material used (dextran, albumin, silicones, and poly(ethylene glycol)). For those systems, the reported hydrodynamic mean diameter ranges from 10 up to 3500 nm.

In recent years, considerable advances have been made in high-resolution in vivo imaging methods for monitoring specific molecular or cellular processes. For this purpose and due to the limited number of targets, high relaxivity values and high specificity of the CAs are necessary. When using Gd-based contrast agents, a large number of paramagnetic centres (high concentration of metal) are often needed to reach a sufficient detection level[10-12]. Owing to the large number of iron atoms per crystal, superparamagnetic nanoparticles of iron oxide have strong $T_2$ and $T_2$* effects and therefore can be useful to image low concentrations of a specific molecular process[13-14]. In spite of their widespread applications, inherent problems of iron-oxide systems remain unsolved or difficult to address. The most prominent ones are a broad particle size distribution, lack of crystal phase identification, low value of the saturation magnetisation of the nanoparticles as compared to the bulk materials, particle aggregation in a magnetic field and particle sedimentation in physiological media. Moreover, as the negative contrast agents induce signal voids (darkening) in the image, a fundamental drawback is that the agent cannot be distinguished in a naturally dark area. A new method for imaging, recently reported by Cunningham et al.[15], enables the use of iron oxide nanoparticles as $T_1$-contrast agents. In such context, very small particles are needed (5-10 nm).

The aim of this work is to investigate the use of the ultrasmall maghemite nanoparticles with very narrow size distribution and high magnetisation as MR $T_1$-contrast agents. Here, we report on the synthesis of the nanoparticles, based on the decomposition of $Fe(CO)_5$ in hydrocarbon solvents[16], as well as its thorough physico-chemical characterisation. In order to stabilise the particles in a physiological environment, they have been transferred to an aqueous medium using an ammonium salt and stabilized with addition of sodium citrate at neutral pH. The nanoparticles remained quite monodisperse, as characterised by TEM and XRD, as well as superparamagnetic at room temperature, as determined by magnetic measurements. The relaxometric measurements performed suggest the potential use of this system as a $T_1$-contrast agent due to the high magnetisation of the nanoparticles at clinical field strengths. After functionalisation with specific biological ligands, these iron oxide nanoparticles could be used as molecular imaging probes[17].

## Experimental

### Materials

Iron pentacarbonyl ($Fe(CO)_5$, 99.999%), oleic acid (99%), dioctyl ether (99%), tetramethylammonium hydroxide (TMAOH, 25 wt. % in $H_2O$), sodium citrate, potassium dichromate ($K_2Cr_2O_7$, 1/25 M in $H_2O$), tin (II) chloride ($SnCl_2$, 98%), mercury (II) chloride ($HgCl_2$, 99.5%), sodium diphenylamine-4-sulfonate and chloric acid (HCl, 37%) were purchased from Sigma Aldrich (Spain) and used as received without further purification.

### Synthesis of hexane-dispersed maghemite nanoparticles

Maghemite nanoparticles were synthesised by decomposition of iron pentacarbonyl in dioctyl ether, in the presence of oleic acid, a non-polar surfactant. Briefly, 0.32 ml (1 mmol) of oleic acid were added to 20 ml of dioctyl ether (solvent) and heated under argon atmosphere up to 90ºC. Then, 0.1 ml (0.75 mmol) of $Fe(CO)_5$ was added to the reaction mixture. The temperature was raised to 300ºC and kept constant for 30 minutes. Afterwards, the mixture was allowed to cool to room temperature. In order to isolate the nanoparticles from the reaction medium, 40 ml ethanol were added to precipitate the particles. The mixture was centrifuged and the black precipitate was recovered in 20 ml of hexane, followed by another centrifugation to allow the precipitation of any insoluble impurities. The solution was recovered and stored tightly closed and hereafter is referred to as $Fe_2O_3$-hexane.

**Stabilisation of maghemite nanoparticles at physiological pH**

To stabilise the particles in an aqueous solution the organic non-polar surfactant was displaced by the electrolyte, tetramethylammonium hydroxide (TMAOH). 5 ml of an aqueous solution of TMAOH (1.71 wt% in water) were added to 5 ml of the hexane iron oxide dispersion. The mixture was stirred during 12 hours. The nanoparticles were recovered by precipitation with acetone, followed by centrifugation. The supernatant was then discarded, and the precipitate was resuspended in 0.02 ml TMAOH and water was added to bring the total volume to 5 ml. The material was stored and hereafter is referred to as $Fe_2O_3$-water. After addition of 0.02 ml of TMAOH (25 wt% in water) and 8 mg (0.03 mmol) of sodium citrate to a 0.5 ml of $Fe_2O_3$-water solution, the pH was brought to neutral pH by adding 0.1 M $HNO_3$ dropwise. The sample then got clear and stable and hereafter is referred to as $Fe_2O_3$-citrate.

**Characterisation of magnetic nanoparticles**

**Transmission electron microscopy (TEM) and electron diffraction.**

A JEOL JEM-1210 Electron Microscope, operating at 120 keV, was used for the electron diffraction analysis and the transmission electron microscopy. The samples for electron microscopy were prepared by deposition of a droplet of the nanoparticle solution onto a carbon-coated film supported on a copper grid and allowed to dry.

**X-Ray Diffraction (XRD).**

$Fe_2O_3$-water sample (after freeze-drying) was characterised by X-ray diffraction with a Siemens D5000 X-ray powder diffractometer using a Cu K$\alpha$ incident radiation. XRD patterns were analysed by Rietveld refinement with the MAUD program[18].

**Dynamic light scattering (DLS) and ξ-potential.**

Dynamic light scattering and ξ-potential measurements were performed with a Zetasizer Nano ZS (Malvern Instruments), provided with a He/Ne laser of 633 nm wavelength. The determination of the isoelectric point, IEP, was performed using the MPT-2 Autotitrator, an accessory of the Zetasizer Nano ZS. The samples were further dissolved 20 times before the measurements of the isoelectric point.

**Infrared spectroscopy (IR).**

IR spectra (4000-400 $cm^{-1}$) from KBr discs of $Fe_2O_3$ aqueous solutions (after freeze-drying the aqueous solution) and of $Fe_2O_3$ in hexane solution (after evaporating the solvent) were recorded on a Fourier transform Perkin-Elmer spectrometer.

**Magnetic measurements.**

Hysteresis loops of the $Fe_2O_3$ aqueous solutions at room temperature were performed in a magnetometer VSM-NUVO, MOLSPIN, Newcastle Upon Tyne, UK. For the frozen $Fe_2O_3$-water sample, hysteresis loops at 5 K and the zero field cool-field cool (ZFC-FC) curves were measured with a superconducting quantum interference device (SQUID) magnetometer (Quantum Desing MPMS5XL). The experimental results were corrected for the holder contribution and for a temperature-independent diamagnetic contribution. All the magnetisation data are presented in $Am^2$/kg $Fe_2O_3$.

**Relaxometric measurements.**

The NMRD (Nuclear Magnetic Resonance Dispersion) profiles were recorded from 10 kHz to 10 MHz on a Stelar field cycling relaxometer (Stelar, Mede, Italy). Additional measurements at 20 and 60 MHz were performed on a Bruker Minispec system (Bruker, Karlsruhe, Germany). The stability of the $T_i$ relaxation times was assessed by repeating the $T_i$ measurements of one sample at several times (24, 48 and 36 hours). No relaxivity changes were detected.

**Total iron concentration.**

The iron concentration in 0.2 ml for the $Fe_2O_3$ aqueous solutions was determined by relaxometry measurements at 20 MHz and 330 K after mineralization in acidic conditions (0.6 ml $HNO_3$ and 0.3 ml $H_2O_2$) by microwaves (Milestone MSL-1200, Sorisole, Italy).

The evidence of the $Fe^{2+}$ ion in $Fe_2O_3$-water solution was determined by titration with potassium dichromate ($K_2Cr_2O_7$).

## Results and discussion

### Characterisation of $Fe_2O_3$-hexane nanoparticles

Iron oxide nanoparticles, crystalline and monodisperse, have been synthesised from the decomposition of the $Fe(CO)_5$ in an organic solvent. Oleic acid creates a shell around the inorganic iron oxide core through the carboxylate oxygens resulting in a steric repulsion between the particles thus preventing their aggregation and precipitation. Transmission electron micrograph of $Fe_2O_3$-hexane solution showed rather spherical, monodisperse and well-formed nanocrystals (see Fig. 1A). By measuring more than 260 particles the mean diameter obtained is 4.9 ± 0.7 nm, with a standard deviation of 14% (see Fig. 1B), where 94% of the particles display sizes between 3.5 and 5.75 nm, and none of them were larger than 8 nm. It has to be noted that for very small particle sizes (~ 5 nm) a difference of 0.5 nm in diameter already means a standard deviation of 10%. High crystallinity of the magnetic particles is deduced from the good definition of the electron diffraction ring pattern (see Fig. 1C). It corresponds to a polycrystalline diffraction pattern (the electron beam is larger than a single particle). The diffraction rings can be equally well indexed considering any of the two iron oxide crystal structures: maghemite ($\gamma$-$Fe_2O_3$) or magnetite ($Fe_3O_4$). Complementary analyses have been carried out to discern which one of the structures is present in our sample (results shown in the next section).

**Characterisation and stabilisation of $Fe_2O_3$-water nanoparticles**

Before these particles can act as contrast agents, they must be stabilised in a physiological aqueous environment. The redispersion of $Fe_2O_3$-hexane nanoparticles in water requires the removal of the surfactant layer, oleic acid, and subsequent replacement with TMAOH. Fig. 2 shows a photograph of $Fe_2O_3$-hexane (top of the two-phase mixture, Fig. 2A)) and of $Fe_2O_3$-water (bottom of the two-phase mixture, Fig. 2B); that is, before and after the phase transfer. In both cases the iron concentration was measured, obtaining the same value in hexane and in water solution, pointing to an exchange yield of 100%.

X-ray diffraction was performed to a lyophilised aliquot of the $Fe_2O_3$-water sample. Because of the modest amount of available sample and the small size of the crystalline particles, the peaks of the diffractogram are not very well defined (see supplementary data). They can be indexed either to maghemite or magnetite. The calculated crystallite size, according to the Rietveld refinement of the pattern, was 5 ± 1 nm, both for maghemite and magnetite, a value which is in agreement with the one obtained by TEM (results not shown).

In order to discern if our iron oxide particles are built up of magnetite ($Fe_3O_4 \equiv 2Fe^{3+}Fe^{2+}4O^{2-}$) or maghemite ($Fe_2O_3 \equiv 2Fe^{3+}3O^{2-}$), the existence of $Fe^{2+}$ ions was determined by the titration with potassium dichromate. No significant $Fe^{2+}$ ion concentration was determined, suggesting the maghemite phase as the crystalline phase. Moreover, the solution presents a red colour, as it is expected for maghemite solutions.

Keeping in mind the biological applications, it was necessary to decrease the pH value from basic to neutral pH. For this purpose, nitric acid ($HNO_3$) was used also in the presence of citrate sodium in order to avoid the agglomeration and precipitation of the nanoparticles at pH lower than 8.

TEM images were also performed for $Fe_2O_3$-citrate resulting in the same particle characteristics as before (see Fig. 3A). The mean particle diameter was 4.8 ± 0.6 nm, with a standard deviation of 12%, where 98% of the particles display sizes between 3.25 and 5.75 nm (see Fig. 3B), being the largest particles 6.25 nm in diameter. The electron diffraction pattern (see Fig. 3C) displays the same diffraction rings as before. The results suggest that no physico-chemical changes occurred during the organic solvent-water exchange.

The hydrodynamic diameter, $d_{HYD}$, is a useful measurement that will define the final biological application of the material studied. Dynamic light scattering size measurements were performed for the three samples: $Fe_2O_3$-hexane, $Fe_2O_3$-water and $Fe_2O_3$-citrate, resulting in hydrodynamic diameters of 12 ± 2 nm, 8 ± 2 nm and 18 ± 4 nm, respectively. We interpret the increase in size for the $Fe_2O_3$-citrate sample as due to incorporation of the citrate ligands at the surface of the iron particles that may induce some aggregation between pairs of particles.

Subsequently, the zeta potential ($\xi$) and the isoelectric point (IEP) for the $Fe_2O_3$ aqueous solutions were estimated. The zeta potential value can be used as an indicator of the stability of a colloidal system. The higher the $\xi$ absolute values, the higher the net electrical charge onto the surface of the particles and, therefore, the larger the electrostatic repulsion between particles. The theoretical limit of stability is |30| mV, i.e. a colloidal system will be stable if its zeta potential is higher than 30 mV or lower than -30 mV. The

isoelectric point (IEP) specifies the pH at which the net electrical charge onto the surface of the particles is zero, i.e., the pH of lowest stability of the system[19]. In Fig. 4, the behaviour of the zeta potential for the $Fe_2O_3$-water and for the $Fe_2O_3$-citrate versus the pH is shown. For the $Fe_2O_3$-water, $\xi$ is lower than -30 mV for pH > 7.5, and higher than 30 mV for pH < 4.3, with an isoelectric point of 6.1, suggesting a lack of stability in the range of physiological pH. The addition of citrate shifts the isoelectric point down to 1.7. This system will be stable at pH higher than 4.1. We believe that this is a result of adsorption of citrate anions to the positively charged particle surface, which then enables the stabilisation of the particles at neutral pH.

Finally, a clear evidence of the removal of the organic layer from the surface of the particles was also provided by FTIR spectroscopy measurements of dried samples mixed with KBr. In the case of the hydrophobic particles, oleic acid is observed, while in aqueous solution, for $Fe_2O_3$-water and for $Fe_2O_3$-citrate, the organic ligand has been replaced by TMAOH. Further, the spectrum of the $Fe_2O_3$-citrate shows clearly the adsorption of citrate anions but still conserving some of the TMAOH ligands (see supplementary data).

Magnetic measurements of the $Fe_2O_3$-water sample are shown in Fig. 5. The magnetisation curve at room temperature plotted on Fig. 5 shows that the $Fe_2O_3$-water sample exhibits superparamagnetic behaviour deduced by the zero coercitive field and the zero remanent magnetisation values. At 5 K, the maghemite nanoparticles show the expected ferrimagnetic behaviour with a coercive field of 96 Oe (see lower inset in Fig. 5). The sample is already saturated at 10 kOe, with a saturation magnetisation value ($M_S$) of 74 $Am^2/kg$ $Fe_2O_3$. Data at room temperature were fitted to a Langevin function obtaining the following values: $M_S = 68$ $Am^2/kg$ and $d_{mag} = 5$ nm, results which are in good agreement with TEM measurements, and a magnetisation saturation value only 10% lower than the value for the bulk material (lit.,[20] 76 $Am^2/kg$ $Fe_2O_3$ at 298 K). Interestingly enough, our iron oxide nanoparticles exhibit a strong induced magnetisation, close to the bulk value, even after decreasing their size up to 5 nm contrasting with the low saturation magnetisation reported in many other maghemite particulated systems[21-23].

The zero field cool - field cool (ZFC-FC) curves of the $Fe_2O_3$-water sample (upper inset in Fig. 5) describe the temperature dependence of the magnetisation. The ZFC curve gives information about the ferri-superparamagnetic transition of the system, which occurs at the temperature of the maximum magnetisation value, the blocking temperature ($T_B$). For our system a $T_B$ value of 15 K was obtained. The splitting of the curves just below $T_B$ and the sharp maximum of the ZFC curve once more stand for a very narrow particle size distribution.

No significant changes in the magnetic behaviour were detected when comparing the above described results with the ones obtained for the $Fe_2O_3$-citrate system (see supplementary data). The shape of the magnetisation loop at room temperature for the $Fe_2O_3$-citrate system also points out to a superparamagnetic system with a saturation magnetisation value of 65 $Am^2/kg$ $Fe_2O_3$, confirming the magnetic stability of the sample at different pHs.

Some physical properties of $Fe_2O_3$-water and of $Fe_2O_3$-citrate are summarised in Table 1.

As previously mentioned, MR contrast agents act by shortening the relaxation times of the surrounding protons, $T_1$ and $T_2$, because of their inherent magnetic properties. The effectiveness of a CA is usually expressed as relaxivities, $r_1$ and $r_{2,}$ per mM of metal. Relaxivities of $Fe_2O_3$-water nanoparticles placed in water with TMAOH are shown in Table 2. $r_1$ values are higher than those observed for the paramagnetic complexes, such as Gd-DTPA, due to its high magnetisation, while $r_2$ values are much lower than values observed for the superparamagnetic particles of larger size. The magnitude of $r_1$ is dependent on the magnetisation of the CA, the electron spin relaxation, the size of the magnetic crystal and the accessibility to the CA of bearing nuclear spins of the tissue. The magnitude of $r_2$ reflects the ability of the CA to produce local magnetic inhomogeneities. At high field $r_1$ decreases while $r_2$ increases when the diameter of the crystal is larger. The ratio $r_2/r_1$ is therefore an indicator of the relaxometric properties of the CA and it serves to classify the type of MR CA as $T_1$-CA or $T_2$-CA[6]. In general, for paramagnetic chelates $r_2/r_1$ varies between 1-2 and for the superparamagnetic colloids it can be as large as 50. Because of the small size of the iron oxide particles evaluated, the ratio $r_2/r_1$ corresponds to the typical values expected for paramagnetic complexes.

Relaxivity measurements of the $Fe_2O_3$-citrate were also performed (see Table 3). In comparison with relaxivity values of $Fe_2O_3$-water, the $r_1$ values for $Fe_2O_3$-citrate are similar but the $r_2$ values have increased. This increase of the ratio $r_2/r_1$ can be considered as an indicator of the clustering between particles[24]. This observation matches very well with the hydrodynamic ratio measurements which also indicated pairing of the particles when coated with citrate.

In order to understand and quantify the mechanisms governing the relaxation phenomenon of the system, a nuclear magnetic relaxation dispersion (NMRD) profile was recorded and fitted according to the usual model[25]. The NMRD profile displays $r_1$ over a wide range of magnetic fields and provides the information necessary to asses the properties of a contrast agent at any field strength. It also gives information on the particle diameter (d) and the value of the saturation magnetization (Ms). The relaxivity profile ($r_1$) of $Fe_2O_3$-water nanoparticles is represented in Fig. 6. The relaxivity, $r_1$, of this complex behaves as expected for ultrasmall superparamagnetic particles. The high field inflexion point named dispersion of the longitudinal relaxation profile is roughly given by the condition $\omega_i\tau_D = 1$ in which $\tau_D = r^2/D$ with r being the radius of the crystal and D the relative diffusion coefficient of water with respect to the particle. $\tau_D$ is the diffusion correlation time which defines the time during which the diffusing proton is influenced by the superparamagnetic particle. After the high field dispersion, the longitudinal relaxation decreases rapidly to zero. The NMRD profile was analysed quantitatively according to equation 31 (lit.[25]) and the values obtained from the fitting are summarised in Table 4. Theoretical values obtained from the fitting are in good agreement with those obtained by other techniques, pointing once more to a very narrow distribution of the particle size and showing consistency within all the used characterisation techniques.

## Conclusions

The present investigation focused on the study of new contrast agents for molecular imaging. Ultrasmall iron-oxide nanoparticles with a very narrow particle size distribution and a high saturation magnetisation value have been obtained. They have been stabilised in water at physiological pH and tested as $T_1$ MRI contrast agents. After functionalization with biological ligands these nanoparticles could be used as magnetic reporters for molecular imaging probes.

## Acknowledgements

This work has been partially financed by the MEC (MAT 2003-01052, MAT2006-13572-C02-01 and ESP2002-03862) and by the Generalitat de Catalunya (Project2005SGR00452). Elena Taboada acknowledges the FPU grant from MEC (AP-2004-2447). The program EMIL, NOE, FP6 is also acknowledged.

## Figure Captions.

**Figure 1.** (A) TEM images of $Fe_2O_3$-hexane. (B) Particle size distribution histogram. (C) Selected area electron diffraction (SAED) pattern with indexed planes.

**Figure 2.** Photograph of two-phase mixtures with $Fe_2O_3$ nanoparticles dispersed in hexane (A) and in water (B).

**Figure 3.** (A) TEM images of $Fe_2O_3$-citrate. (B) Particle size distribution histogram. (C) Selected area electron diffraction (SAED) pattern with indexed planes.

**Figure 4.** Behaviour of the Zeta Potential versus pH for the $Fe_2O_3$-water and $Fe_2O_3$-citrate samples.

**Figure 5.** Main panel: magnetisation curves at 298 K (circles) and 5 K (squares) for the $Fe_2O_3$-water sample. Upper inset: ZFC-FC curves. Lower inset: enlargement of the magnetisation loop at 5 K.

**Figure 6.** $^1$H NMRD profile of the $Fe_2O_3$-water sample at 310 K.

**Figures**

**Fig. 1**

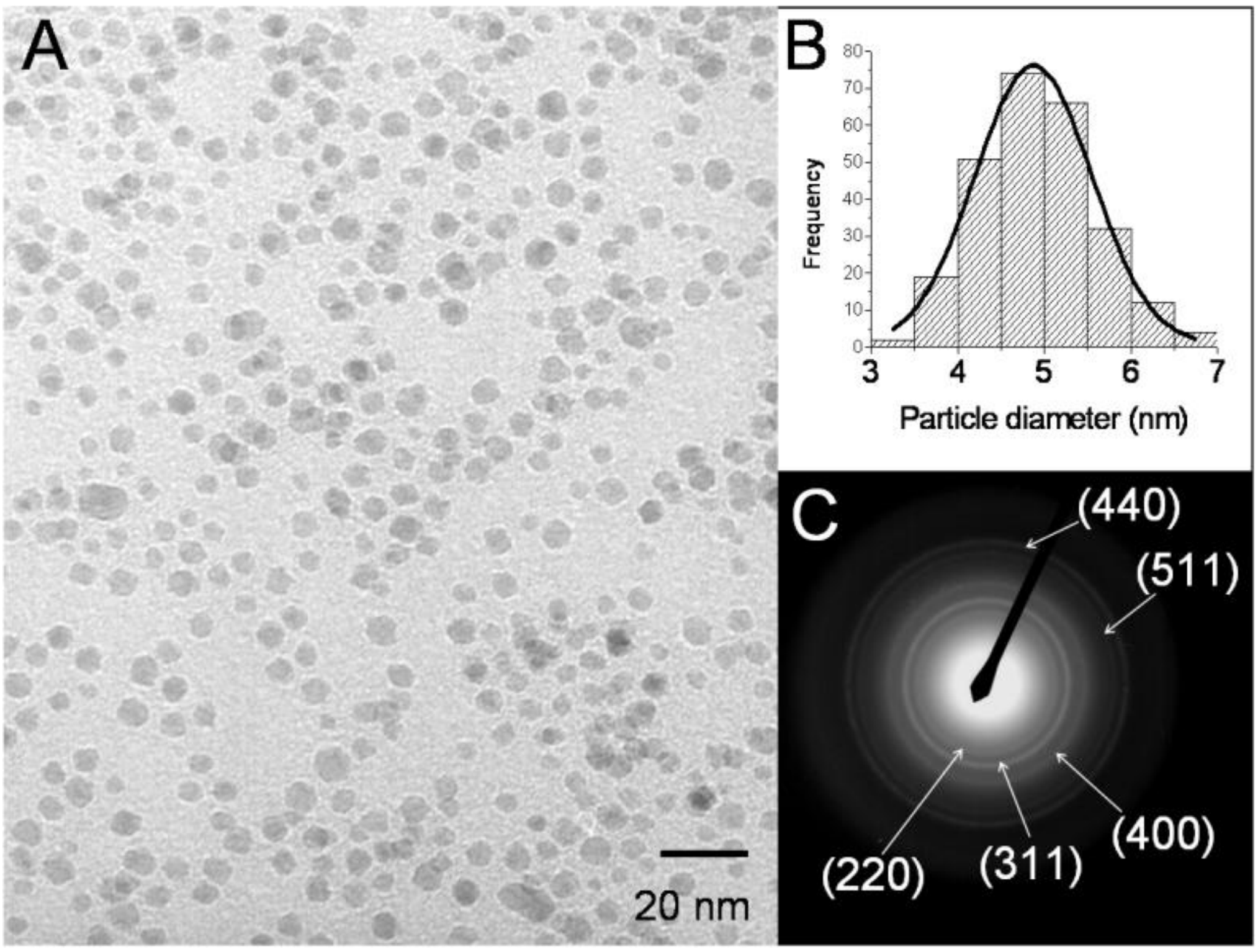

**Fig. 2**

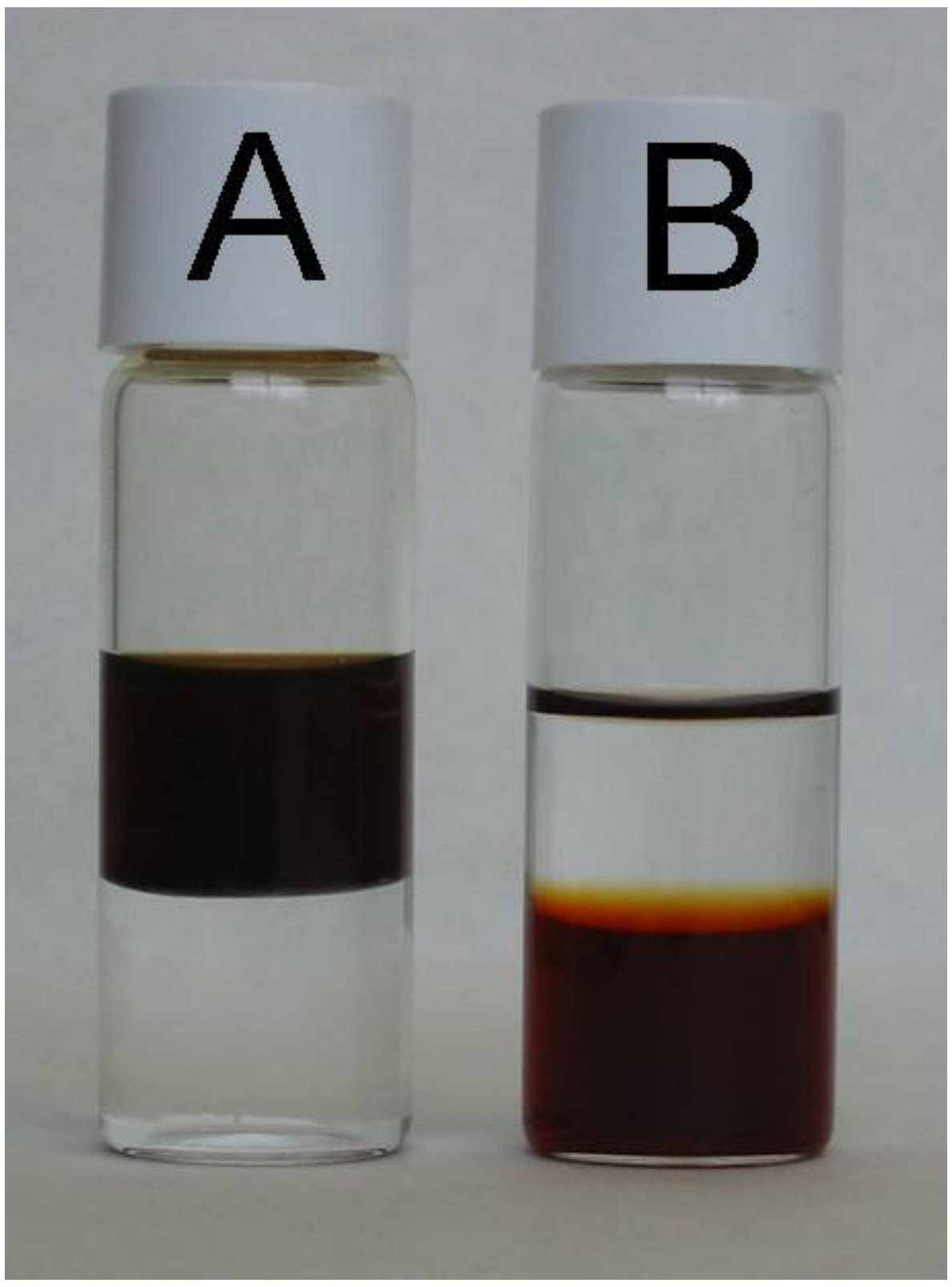

**Fig. 3**

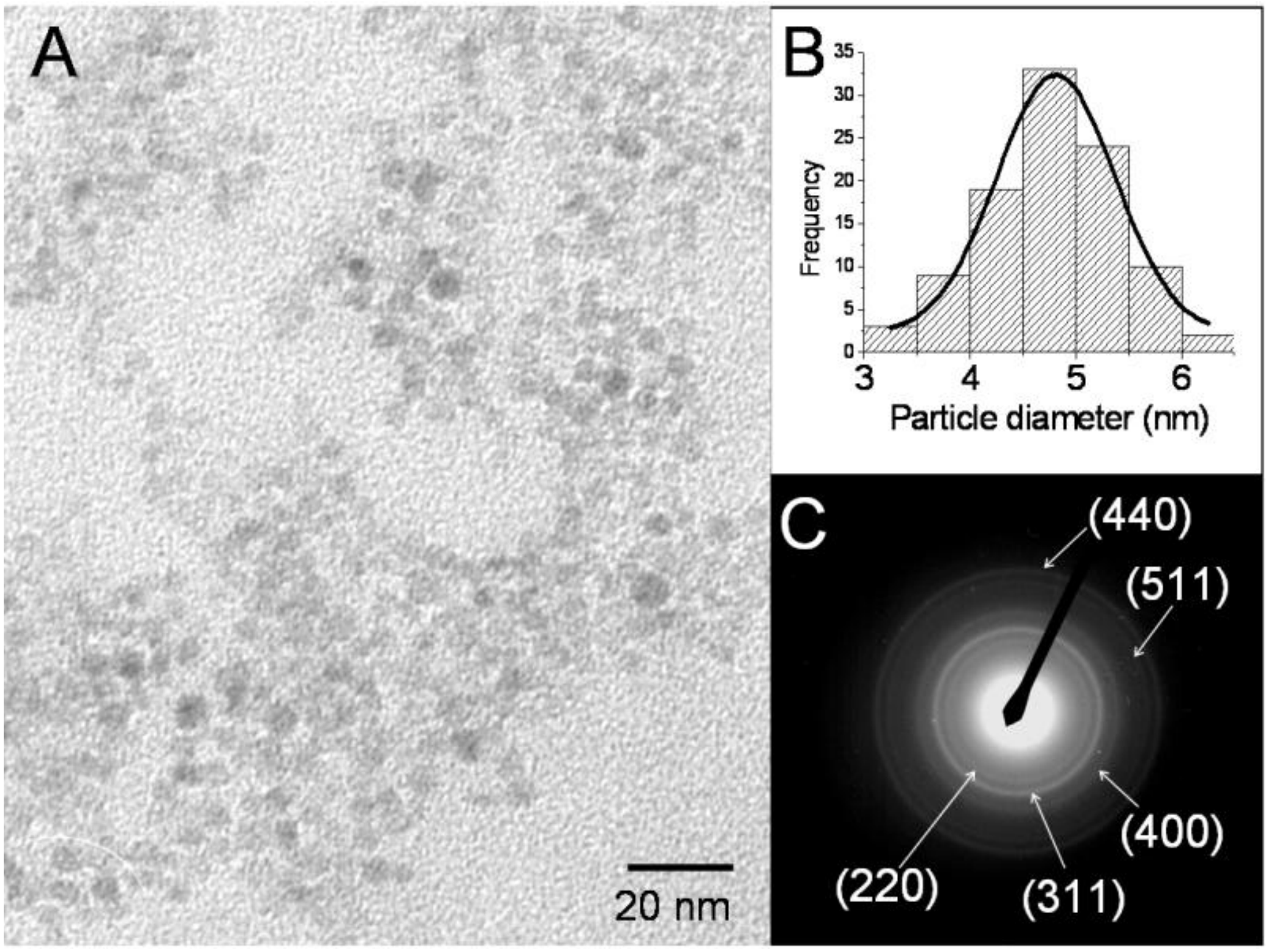

**Fig. 4**

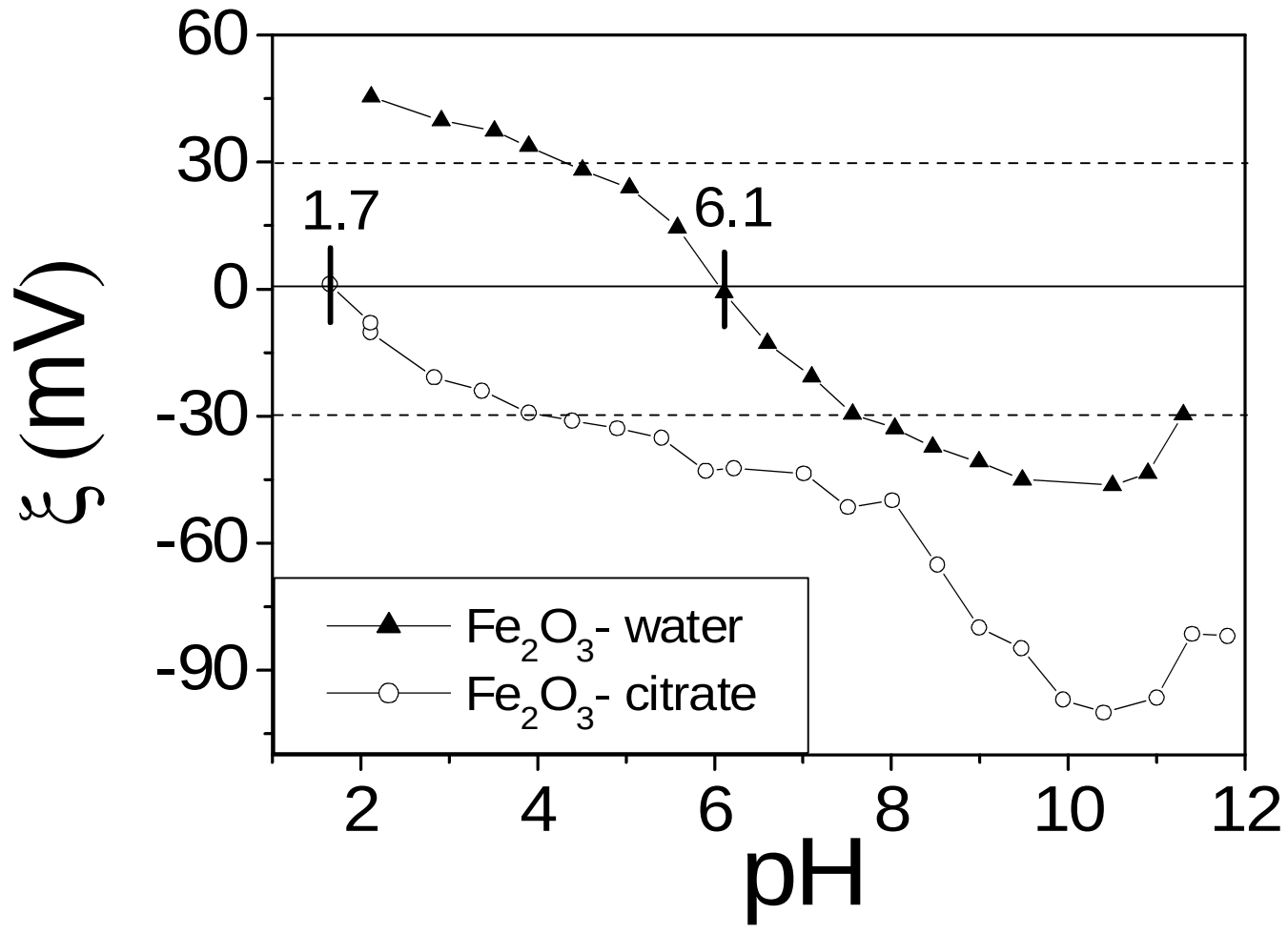

**Fig. 5**

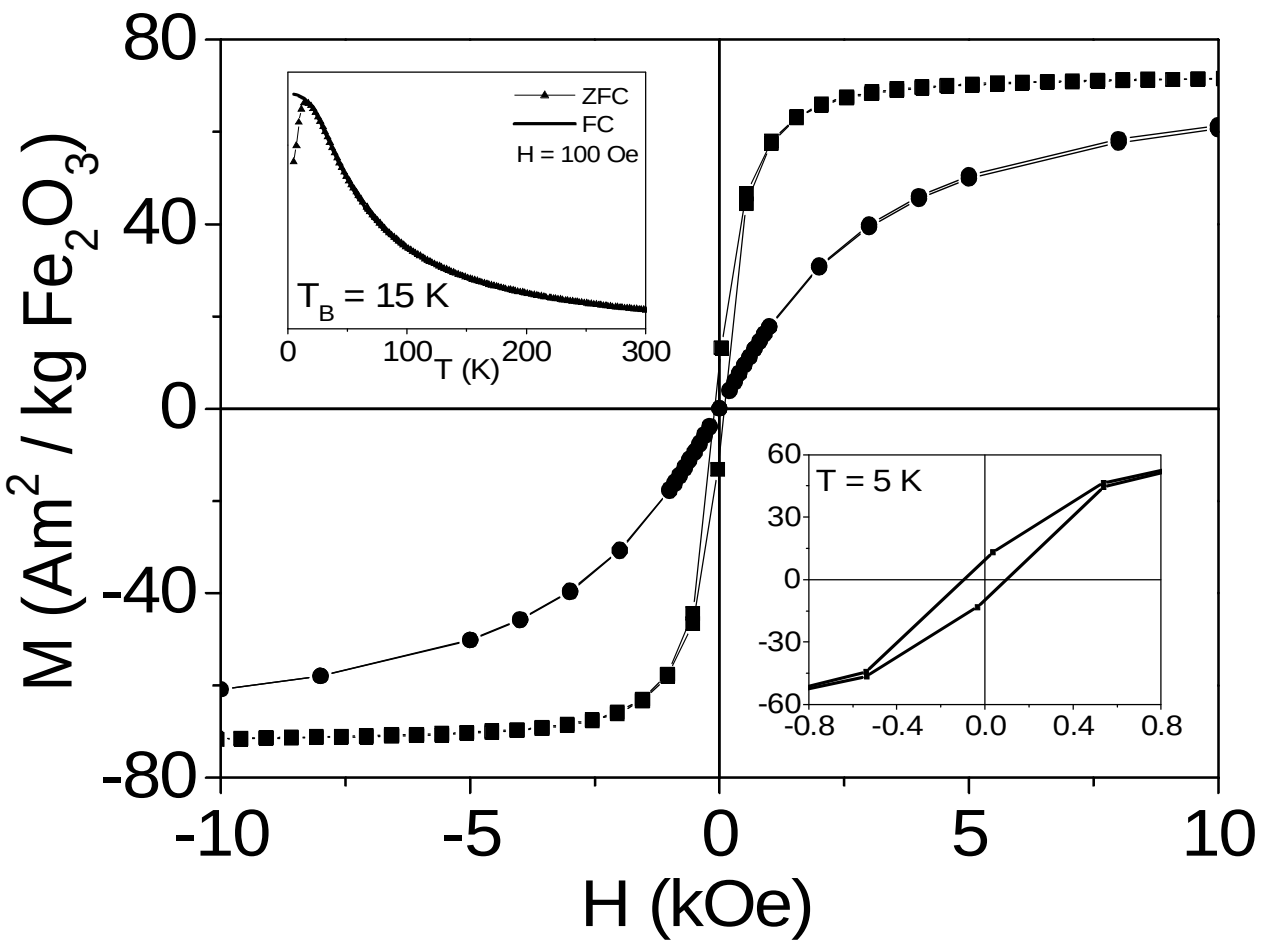

**Fig. 6**

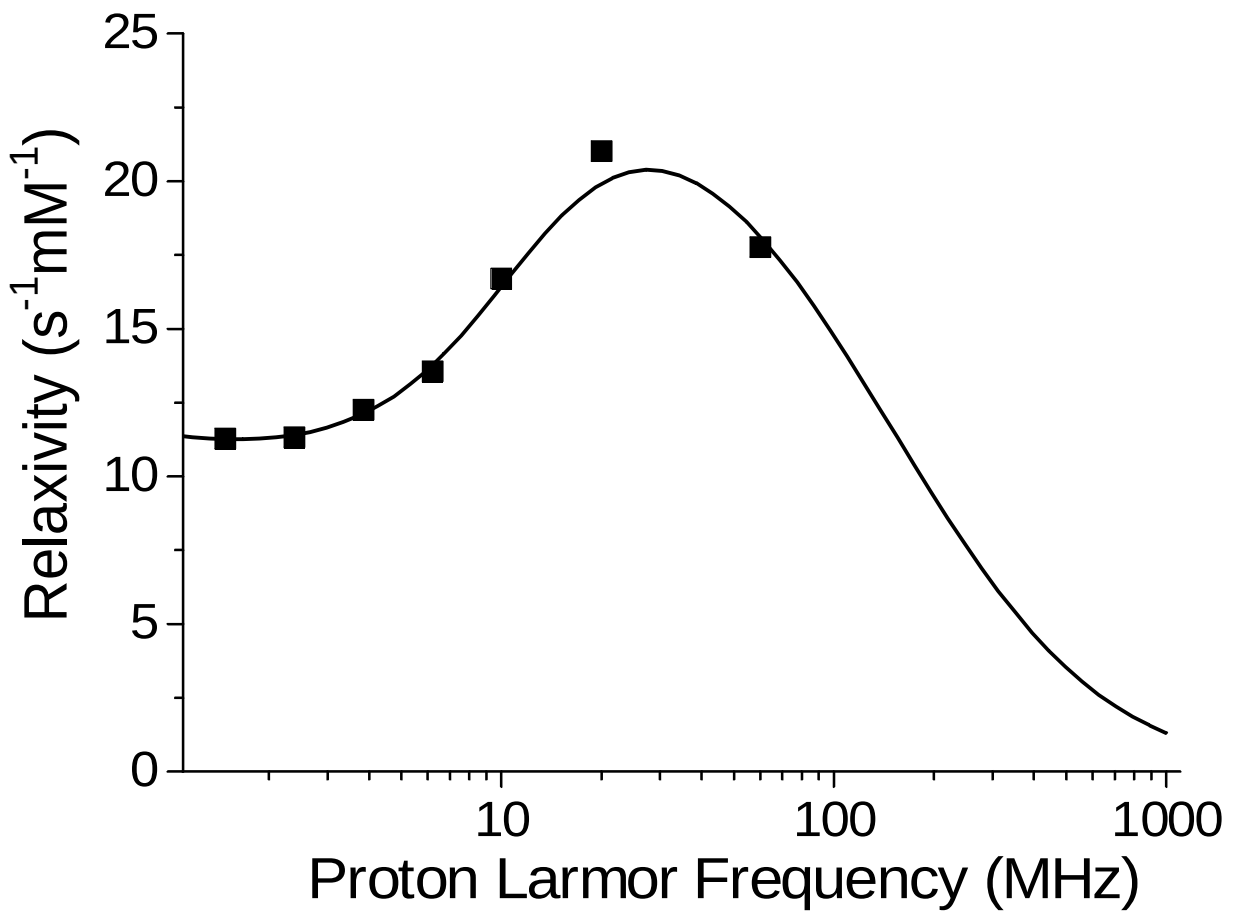

# Tables

## Table 1

Physical properties of $Fe_2O_3$-water and $Fe_2O_3$-citrate solutions

| Physical property | $Fe_2O_3$-water | $Fe_2O_3$-citrate |
|---|---|---|
| $d_{TEM}$ (nm) | 4.9 ± 0.6 | 4.8 ± 0.6 |
| $d_{XRD}$ (nm) | 5 ± 1 | |
| $d_{HYD}$ (nm) | 8 ± 2 | 18 ± 4 |
| Crystalline phase | $\gamma$-$Fe_2O_3$ | $\gamma$-$Fe_2O_3$ |
| Colour of solution | red | red |
| IEP (pH unities) | 6.1 | 1.7 |
| pH | 12.5 | 7.4 |
| Ms ($Am^2/kg$) at 298 K | 68 | 65 |
| Hc (Oe) at 5 K | 96 | 73 |
| $T_B$ (K) | 15 | 11 |

TEM: transmission electronic microscopy; XRD: X-Ray diffraction; HYD: hydrodynamic; IEP: isoelectric point; $M_S$: saturation magnetisation; $H_C$: coercive field; $T_B$: blocking temperature.

**Table 2**

Relaxivity values of $Fe_2O_3$-water solutions at basic pH and 310 K

| | $r_1$ ($mM^{-1}s^{-1}$) | $r_2$ ($mM^{-1}s^{-1}$) | $r_2/r_1$ |
|---|---|---|---|
| 20 MHz | 20.81 | 28.61 | 1.38 |
| 60 MHz | 17.59 | 35.75 | 2.03 |

**Table 3**

Relaxivity values of $Fe_2O_3$-citrate solutions at neutral pH and 310 K

| | $r_1$ ($mM^{-1}s^{-1}$) | $r_2$ ($mM^{-1}s^{-1}$) | $r_2/r_1$ |
|---|---|---|---|
| 20 MHz | 20.76 | 51.02 | 2.46 |
| 60 MHz | 14.50 | 66.90 | 4.61 |

**Table 4**

Parameters obtained from the fittings of the NMRD profile at 310 K for the $Fe_2O_3$-water sample.

| NMRD profile | | TEM | Magnetic measurements | |
|---|---|---|---|---|
| $d^a$ (nm) | $Ms^a$ ($Am^2$/ kg $Fe_2O_3$) | $d^b$ (nm) | $Ms^c$ ($Am^2$/ kg $Fe_2O_3$) | $d^b$ (nm) |
| 5.7 | 73 | 4.9 | 68 | 5 |

[a]diameter and $M_S$ from NMRD profile; [b]diameter from TEM measurements; [c]diameter and $M_S$ from magnetic measurements